\documentstyle[11pt,paspconf,epsf]{article}

\begin{document}

\title{Submillimetre source counts: first results from SCUBA}

\author{Robert G. Mann}
\affil{Astrophysics Group, The Blackett Laboratory,\\
Imperial College of Science, Technology \& 
Medicine, \\ Prince Consort Road, London, SW7
2BZ, United Kingdom}

\author{and The UK Submillimetre Survey Consortium\altaffilmark{1}}
\affil{}

% Notice that some of these authors have alternate affiliations, which
% are identified by the \altaffilmark after each name.  The actual alternate
% affiliation information is typeset in footnotes at the bottom of the
% first page, and the text itself is specified in \altaffiltext commands.
% There is a separate \altaffiltext for each alternate affiliation
% indicated above.

\altaffiltext{1}{A.W. Blain (Cambridge), J.S. Dunlop (Edinburgh), A.N.
Efstathiou (ICSTM), D.H. Hughes (INAOE), R.J. Ivison (UCL),
A. Lawrence (Edinburgh), M.S. Longair (Cambridge), S.J. Oliver (ICSTM), 
J.A. Peacock (Edinburgh), M. Rowan-Robinson (ICSTM),
S.B.G. Serjeant (ICSTM)}

% The abstract is entered in a LaTeX "environment", designated with paired
% \begin{abstract} -- \end{abstract} commands.  Other environments are
% identified by the name in the curly braces.

% Poster authors ONLY may omit the abstract in order to gain a little
% more page space for the text of the poster.

\begin{abstract}
The SCUBA submillimetre camera has opened up new
possibilities for 
tracing the evolution of active star formation in dusty galaxies
 to high redshift, with profound implications for our
understanding of the star formation history of the Universe and
the contribution from galaxies
to the anisotropy of the microwave sky. We review results from several
submillimetre surveys started during SCUBA's first year of operation,
and discuss their future development, together with other projects
that will greatly improve our understanding of the extragalactic
point source
contribution to the submillimetre sky in the era of MAP and {\em Planck\/}.

\end{abstract}

% Keywords should be included, but they are not printed in the hardcopy.


\setcounter{footnote}{1}

% That's it for the front matter.  On to the main body of the paper.
% We'll only put in tutorial remarks at the beginning of each section
% so you can see entire sections together.

\section{Introduction}

Advances in modern observational astronomy are usually led by innovation in
detector technology, rather than through inspiration on the part of
observers. A good example of this is the recent upsurge of interest in
submillimetre astronomy, following the 
installation of the Submillimetre Common User Bolometer Array 
(SCUBA\footnote{SCUBA Home Page: {\tt www.jach.hawaii.edu/JCMT/scuba}}:
Holland et al. 1999) on the James Clerk Maxwell Telescope  on
Mauna Kea, Hawaii. The (sub)millimetre is one of the few regions of
the electromagnetic spectrum remaining largely unexplored by modern
astronomy, and, as we detail below, SCUBA is the first instrument to
cross the sensitivity threshold beyond which serious extragalactic
astronomy in the submillimetre becomes not only possible, but very
powerful. 
This has led several groups to start survey programmes
with SCUBA, and this article reviews their progress to date. 
These surveys impact on the study of CBR foregrounds in three ways:

\begin{enumerate}

\item the central wavelength of the most sensitive SCUBA band
(850$\mu$m) matches that of one of the channels of the
{\em Planck High Frequency Instrument\/} (HFI: Puget et al. 1998), so
SCUBA is directly sampling one of the foregrounds to be seen by
Planck, albeit at fainter flux densities
than HFI will reach

\item the source counts resulting from the SCUBA surveys will provide
important constraints on the models of galaxy evolution required to
predict the level of point source contamination in microwave sky maps
produced by CBR experiments in less directly studied passbands 

\item SCUBA observations may resolve the cosmic infrared/submm
background (Puget et al. 1996) detected in COBE maps.

\end{enumerate}

These issues are addressed more directly in the contributions to this
volume by Eric Gawiser, Bruno Guiderdoni, Luigi
Toffolatti,  and collaborators, so
we shall concentrate here on observational aspects of SCUBA survey
programmes.

\section{Why study galaxies in the submillimetre?}

IRAS revealed a strongly evolving population of starburst
galaxies forming stars at rates of a few hundred solar masses per year
in heavily obscured regions, reminiscent of
the formation of massive stars in giant molecular clouds in the
Galaxy. Typical starbursts emit $\sim50$\% of their bolometric
luminosity in the rest-frame far--infrared, as the 
UV radiation from massive young stars heats the dust around them, while
this fraction can increase to $\sim 90$\% in ultraluminous infrared
galaxies (ULIRGs: those with $L_{\rm FIR} > 10^{12}L_{\odot}$),
although some of these are partially powered by active galactic nuclei
(Genzel et al. 1998). The absorption of such a large fraction of the
UV/optical radiation from massive stars in starburst galaxies 
necessarily means that studies performed in those bands will
underestimate the star formation rates in individual starbursts 
(as indicated by Cram et al. 1998, from the comparison of estimates of
the rates of star formation  in a sample of $\sim 700$ local galaxies
derived from U band, H$\alpha$, far--infared and decimetric radio
continuum luminosity\footnote{While these and most other conventional 
methods of estimating the star formation rate of a galaxy determine
the luminosity from the massive stars it contains, the exact stellar
mass range probed by each method is slightly different, and the conversions
from luminosity density to star formation density required to put each
point on the ``Madau diagram'' are uncertain to factors of a few
(e.g. Rowan--Robinson et al. 1997) due to differing assumptions about
the stellar initial mass function.}), as well as allowing the
possibility of the complete omission of a population of
heavily--obscured starbursts from the cosmic star formation census. Since
most massive stars are made in starburst galaxies this can have a
significant effect on the global star formation rate, and, once
extinction corrections (Meurer et al. 1997, Pettini et
al. 1998) are applied, the
star formation history of the Universe deduced from UV/optical studies
can change substantially (e.g. Steidel et al. 1999) from the
well--known shape of the much--touted ``Madau diagram'' (Lilly et al.
1996, Madau et al. 1996).

\begin{figure}
\plotfiddle{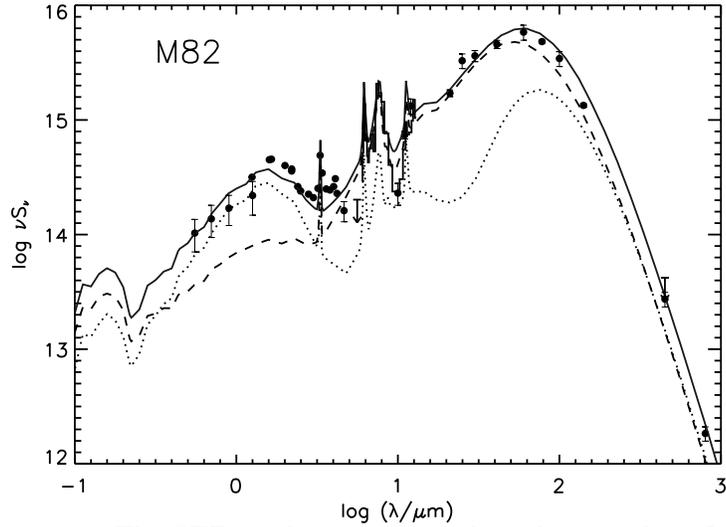}{6.2cm}{0}{60}{60}{-200}{-235}
\caption{The SED of the prototypical starburst
galaxy M82 (from Efstathiou et al. 1999: ERRS). Observational data
points are plotted, together with model
SEDs resulting from two bursts of star formation (dotted and dashed
lines), whose sum (solid line) provides a good fit to the galaxy's
SED over the range 0.5--850 $\mu$m: see ERRS for details.}
\end{figure}

\begin{figure}
\plotfiddle{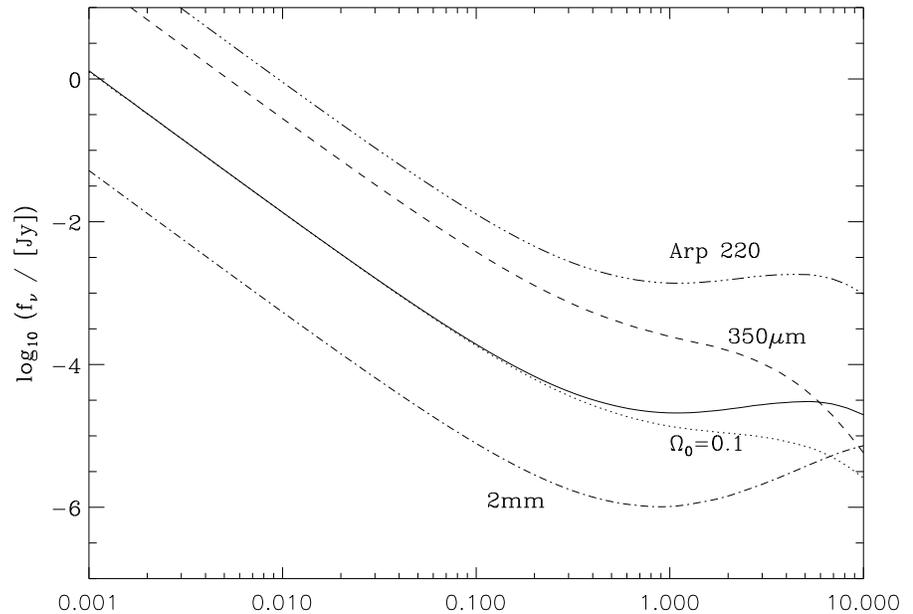}{7.2cm}{90}{52}{52}{200}{-25}
\caption{Starburst submillimetre flux density as a function of
$z$. The solid (dotted) line shows the result of taking 
M82 (the solid line of Fig. 1) and observing
it at 850$\mu$m at increasingly high redshift in a Universe with 
$\Omega_0=1$ ($\Omega_0=0.1$) and $\Lambda=0$. The dashed  
(dot--dashed) lines show the results of doing this at
350$\mu$m (2mm), while the
 dot--dot--dot--dashed line shows the effect on the solid line of 
replacing M82 ($L \sim 10^{10} L_{\odot}$) with Arp 220, which is
about $\sim 60$ times more
luminous, and has a slightly different SED (A. Efstathiou, priv.
comm.).}
\end{figure}

In Figure 1 we plot the spectral energy distribution (SED) of the
prototypical local starburst galaxy M82, which shows the characteristic
starburst shape of a peak at $\sim 60$ $\mu$m, and a steep decline
to longer wavelengths, well fitted by a single temperature greybody
curve with emissivity $\propto \nu^{1-2}$ and temperature $\sim 50$ K
(although the physics of the dust emission is certain to be more
complicated than this simple model might seem to imply). The steepness
of this decline means that such galaxies have high negative
$k$-corrections in the submillimetre -- i.e. as they are observed in a
fixed passband at increasingly high redshift, the rest--frame
wavelength of the observed emission moves towards the peak and so the
rest--frame flux increases -- which counterbalance the inverse square law
effect due to the increasing distance, with the result that the
observed submillimetre flux (in a fixed passband) of a
starburst galaxy of a given luminosity is almost independent of
redshift for $1 \leq z \leq 10$ or so (for $\Omega_0 =1$ -- the effect depends
on cosmology through the luminosity distance), as shown in Figure 2.

\section{The Submillimetre Common User Bolometer Array (SCUBA)}

SCUBA  consists (Holland et al. 1998) of two arrays
of bolometers (37 in the long wave array, 91 in the short wave
array) allowing simultaneous imaging in pairs of submillimetre
bands (the most sensitive being 850$\mu$m and 450$\mu$m) over
a field of view of $\sim2.3$ arcmin diameter, or photometry at
1.1, 1.3 or 2.0 mm using single detectors in each band. The bolometer
feedhorns are close--packed, but
finite gaps between them necessitate 
a series of offset observations (performed by ``jiggling'' the
secondary mirror of the JCMT) to sample the field of view fully. A
total of 64 pointings is required to produce fully sampled maps at
both 850 and 450 $\mu$m, and these have diffraction--limited angular
resolutions of 14.7 and 7.5 arcseconds (FWHM),
respectively. Each bolometer is $\sim100$ times more sensitive than
the previous JCMT single--element broadband photometer (UKT14) which,
combined with its large number of pixels, means
that SCUBA reaches the plateau in the flux--redshift plot for luminous
starbursts in a sufficiently short time  and large field of view to
make large submillimetre surveys feasible for the first time.

\section{Submillimetre surveys with SCUBA}

Four groups have reported results from SCUBA surveys, and several
others are underway. Two complementary
strategies have been adopted: (i) simple mapping of blank
fields; and (ii) pointed observations
of massive clusters of galaxies, using their gravitational lensing
effect to probe deeper than otherwise possible in a given integration
time. Harnassing gravitational lensing
amplification probes deeper (simplifying follow-up observations in other
wavebands, as well as initial source detection with SCUBA) but it
can introduce into quantitative results uncertainty resulting from
 imprecision in the cluster lens model, and, given the large
beam size, there is the possibility of misidentifying emission from cold
dust around the central galaxy in the cluster core -- as seen by
IRAS in some clusters 
(Bregman, McNamara \& O'Connell 1990)
and as expected in cooling flow models (see Fabian 1994 for a review) --
with emission from a gravitationally--lensed background galaxy along
the line of sight.

\subsection{The SCUBA Lens Survey}

The first SCUBA survey results were reported by Smail, Ivison \&
Blain (1997) who detected six sources in observations of the two
clusters A370 and Cl 2244-02 at $z=0.37$ and 0.33
respectively. They deduced a source density of $(2.4\pm1.0) \times
10^3$ deg$^{-2}$ at a 50\% completeness limit of $\sim4$mJy at 850
$\mu$m, requiring strong evolution of the starburst
population at $z>1$ (on the basis of the IRAS 60 $\mu$m luminosity
function of Saunders et al. 1990, and the galaxy SED models of
Blain \& Longair 1996). The SCUBA Lens survey is now complete
(Smail et al. 1999), with 17 sources detected at
3$\sigma$ or better, from observations of seven clusters over the
redshift range \mbox{$0.19 \leq z \leq 0.41$}, covering a total area of
0.01 deg$^2$ to a depth of 2mJy. Notable amongst the follow-up
of the sources is the study 
of the galaxy SMM J02399-0136 (Ivison et al. 1999) detected in the
field of A370 and identified with a $z=2.8$ galaxy.
It exhibits a hybrid of features associated with
star formation (high molecular gas mass deduced from
CO emission, high radio and H$\alpha$ luminosities) at a rate of
several thousand solar masses per year, and with the presence of
an active nucleus (optical emission line properties and radio
morphology): this mix of features is also seen in some
hyperluminous infrared galaxies (e.g. F10214+4724: Lawrence et
al. 1994). 

\subsection{The UK Submillimetre Survey}

Our own UK Submillimetre Survey comprises two separate blank field  programmes
-- a narrow, deep survey and a wide, shallow 
survey -- which will combine to produce a dataset capable of tracing
the form of the evolution of the starburst population to high
redshift. We have completed our narrow, deep survey (Hughes et al.
1998) which was undertaken in the Hubble Deep Field (HDF: Williams et
al. 1996) during a period of exceptionally good weather conditions. We
covered an area of $\sim5.5$ arcmin$^2$ to a 1$\sigma$
noise level of 0.45 mJy at 850 $\mu$m, putting us beyond the classical
confusion limit, given the size of the SCUBA beam. This complicated
the identification of discrete sources, but, on the basis of
simulations (see Hughes et al. 1998 for details), we conservatively
accepted as real and discrete a sample of 5 sources with fluxes
above 2 mJy.
 
The principal reason for undertaking our narrow, deep survey in the
HDF was the existence of a wealth of data in other wavebands in that
field, which would help with the process of identification of our
SCUBA sources. For our five sources we found
four plausible associations, with galaxies at redshifts (spectroscopic
or photometric) $z \simeq$ 1, 2, 3 and 3, while the fifth source had
two candidate counterparts, one with a photometric redshift of $z
\simeq 4$ and the second a faint galaxy lying in the Hubble Flanking
Fields and therefore not amenable to photometric redshift estimation.
Note that the optical/far--infrared flux ratios typical of starburst
galaxies do not preclude our SCUBA sources being high--redshift
galaxies too faint to be detected in the optical HDF, despite its
unprecedented depth, and also that only our brightest source (with a
flux of 7 mJy at 850 $\mu$m) is detectable at 1.3 mm with the IRAM
interferometer, which would yield a more accurate source position,
simplifying its identification.

\subsection{The Hawaii survey}

In the same issue of {\em Nature\/} in which we reported on our HDF
survey, Barger et al. (1998) presented results of pointings in two
blank fields: one in the Lockman Hole, and one in Hawaii deep survey
field SSA13. They detected two sources above 3 mJy,
the brightest of which they deduce to lie in the range $1.5 \leq z
\leq 3.5$.

\subsection{The Canada--UK Deep Submillimetre Survey}

A second wide--area survey is being undertaken in two of the fields of
the Canada--France Redshift Survey (CFRS), which total 200
arcmin$^2$. Results from the first 22 arcmin$^2$ have been presented by
Eales et al. (1999), who report detections of 12 sources with $S_{850
\mu{\rm m}} > 2.8$ mJy. Lilly et al. (1999) found within the
CFRS associations for six of these sources, while a further two have
tentative identifications to the CFRS photometric limit of $I_{\rm
AB}=25$: of these, four have spectroscopic
or photometric redshifts below $z=1$, four lie in the interval $1 \leq z
\leq 3$, while the four unidentifed sources are assumed to have $z>3$.

\section{Summary}

\begin{figure}
\plotfiddle{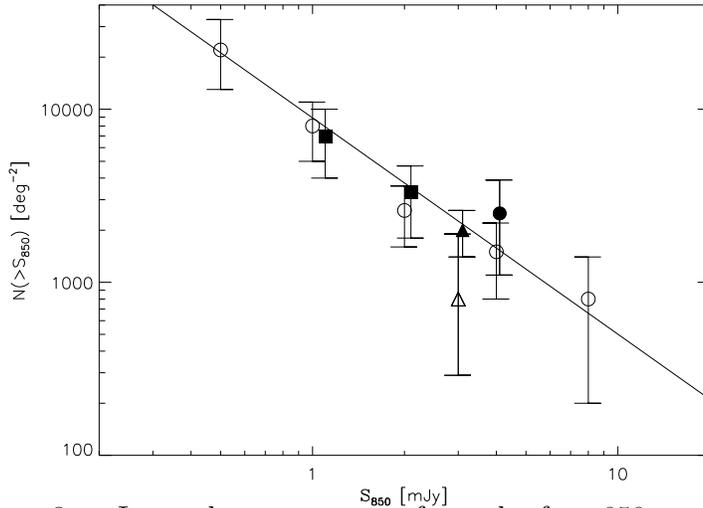}{6cm}{90}{41}{41}{150}{-20}
\caption{Integral source counts from the four 850$\mu$m surveys described in
the text: empty circles (Smail et al. 1999); filled circle
(Smail et al. 1997); empty triangle (Barger et al. 1998); filled
triangle (Eales et al. 1999); and filled squares (Hughes et al.
1998). For clarity, filled symbols are plotted 0.1mJy brighter
than their true flux limit: the line -- $N(>S)=10^4 (S/1{\rm 
mJy})^{-5/4}[{\rm deg}^{-2}]$ -- is not a fit to the data. }
\end{figure}

The results reported to date appear to present a fairly consistent
picture. Figure 3 shows that the integral source count distribution
is already becoming constrained over more than an order
of magnitude in flux density, while Lilly et al. (1999) note that
the redshift distribution they derive appears broadly consistent (given the
small numbers of sources, and the uncertainty in estimating
photometric redshifts at $z>1$) with that of our HDF sources, and that
the $I_{\rm AB}$ band magnitude distributions of the galaxies
associated with the CFRS and SCUBA Lens Survey sources also appear
to agree well. All four groups broadly ascribe their results to a
population of luminous starburst galaxies, which resolve most (perhaps
nearly all?) of the cosmic far--infrared/submillimetre background, have
SEDs broadly similar to that of Arp 220, and are observed to be
forming stars at rates
of several hundred solar masses per year. Much of this activity is
taking place at redshifts where previous UV/optical studies 
(Madau et al. 1996) suggested there should be little star
formation, although uncertainty over the local template to use for
these galaxies means (Eales et al. 1999) that it is not 
clear whether these two views of the star formation history really are 
inconsistent. The revision upwards of the UV/optical star
formation rates at $z>2$ (Steidel et al. 1999) does, however,
make a consistent picture
emerge, in which luminous, dusty starbursts at high
$z$ produce a large fraction of the Universe's stellar mass.

It is tempting to identify these objects as the long--sought
precursors of today's luminous elliptical galaxies, and perhaps this
will become clear by the time these surveys are
complete. The two wide area surveys described here should detect 
a few hundred sources in total, sufficient to
characterise the nature and evolution of this population well
(although optical spectroscopy of most of these objects will not be 
possible, even with 10m class telescopes). In
particular, our UK Submillimetre Survey has been awarded the time
needed to complete our shallow ($3\sigma$ depth
of 8 mJy at 850$\mu$m) survey of 640 arcmin$^2$ in three fields with
deep ISOPHOT 175 $\mu$m data: these sources should be bright enough to
follow up at 1.3mm with the IRAM interferometer, to produce more
precise positions, making the association procedure easier, while
non--detection of any sources in the deep 175$\mu$m data robustly
indicate that the galaxy lies at $z>2$.

As discussed in other contributions to this volume, the SCUBA results
described here are already providing information on the extragalactic
point source population that will be important to MAP, while, by the
time {\em Planck\/} is launched in 2007, there will be several other
very powerful facilities available for studying galaxies in the
(sub)millimetre. Firstly, SCUBA itself may be upgraded\footnote{see
{\tt www.jach.hawaii.edu/JCMT/scuba/upgrades/up.html}}, with suggested
improvements capable of increasing sensitivity so that the confusion
limit at 850$\mu$m can be reached in 2--3 hours of good weather,
rather than $\sim$20 as at present. The first of the completely
new facilities is the Large Millimeter Telescope
(LMT\footnote{LMT Home Page: {\tt www-lmt.phast.umass.edu/}}),
a 50m single dish telescope to be built by a US/Mexican consortium
at a site in Mexico at an altitude of 4,600m. The LMT will operate
over the range 0.85--3.4mm, and its  Bolocam broad band camera should reach
a 10$\sigma$ sensitivity of 0.3 mJy in 1 hour at 1mm (D. Hughes, priv. comm.).
Most important is the planned Large Millimetre Array (LMA), which is likely
to result from a merger of ESO's Large Southern Array\footnote{LSA
Home Page: {\tt puppis.ls.eso.org/lsa/lsahome.html}} project and the
Millimeter Array\footnote{MMA Home Page: {\tt www.mma.nrao.edu}}
planned by NRAO for the same exceptional site at Llano de Chajnantor
in Chile. The design of the LMA is still under discussion, but it will
provide the possibility of sub-arcsec resolution (sub)millimetre
imaging by interferometry to very faint levels, albeit in a small
field of view. This would allow detailed follow up of sources found
in surveys made with SCUBA, LMT,
FIRST\footnote{FIRST Home Page: {\tt astro.estec.esa.nl/First/first.html}}, etc, and also with the HFI on {\em
Planck\/} itself. It is clear that the continued rapid development of
(sub)millimetre astronomy over the next decade will provide the
greatly improved understanding of the extragalactic point source
population required for the success of future CBR experiments like MAP and {\em
Planck\/} as well as facilitating important secondary science 
with these missions, studying the astrophysics of the point sources they
will remove from their maps of the microwave sky.

% Finally, we have a little acknowledgements section.

\acknowledgments
RGM acknowledges support from PPARC and thanks Max
Tegmark and Angelica de Oliveira--Costa for organising an excellent
workshop, and Andreas Efstathiou for help with Figs. 1 and 2. 
The James Clerk Maxwell Telescope is operated by The Joint Astronomy Centre on behalf of the Particle Physics and Astronomy Research
Council of the United Kingdom, the Netherlands Organisation for
Scientific Research, and the National Research Council of Canada. The
principal credit for all the results described here lies with the
SCUBA development team, for creating such a powerful intrument.

% That's the end of the main body of the paper.  Now we will have some
% back matter.

% That's all, folks.
%
% The technique of segregating major semantic components of the document
% within "environments" is a very good one, but you as an author have to
% come up with a way of making sure each \begin{whatzit} has a corresponding
% \end{whatzit}.  If you miss one, LaTeX will probably complain a great
% deal during the composition of the document.  Occasionally, you get away
% with it right up to the \end{document}, in which case, you will see
% "\begin{whatzit} ended by \end{document}".

\end{document}